\documentclass[a4paper,11pt]{article}
\usepackage[utf8]{inputenc}
\usepackage[top=0.95in, bottom=0.75in, left=1.28in, right=0.8in]{geometry}
\usepackage{mathtools,hyperref}
\usepackage{graphics}

\title{\large \textbf {Quadrature Squeezing in the Cavity Mode Driven by Coherent Light and Interacting with Two-Level Atom}}
\author{\large Beyene Bashu and Fesseha Kassahun\\Department of Physics, Addis Ababa University\\P. O. Box 1176, Addis Ababa, Ethiopia}
\begin{document}

\maketitle

\begin{abstract}
We have analyzed the quadrature squeezing in a cavity mode driven by coherent light and interacting with a
two-level atom. We have found that the cavity mode is in a squeezed state, with the maximum quadrature squeezing being
$50\%$ below the vacuum-state level. We have also considered the superposition of a pair of cavity modes,
each driven by coherent light and interacting with a two-level atom. We have found that the superposed cavity modes
are in a squeezed state and the squeezing occurs in both quadratures. We have established that the uncertainty relation
perfectly holds for this case. In addition, we have observed that the sum of the maximum squeezing in the plus and minus
quadratures is $50\%$.
\end{abstract}
\providecommand{\keywords}[1]{\textbf{{keywords:}} #1}
\section{Introduction}
There has been a considerable interest in the analysis of the quantum properties of the squeezed light generated by various 
quantum optical systems [1--9]. A light mode is said to be in a squeezed state if the quantum noise in at least one quadrature 
is below the vacuum-state level, with the product of the uncertainties in the two quadratures satisfying the pertinent uncertainty 
relation. In addition to exhibiting a nonclassical feature, squeezed light has potential applications in precision measurements 
and noiseless communications [10, 11]. It has been established theoretically that a subharmonic generator, a three-level laser 
pumped by electron bombardment, and a three-level laser pumped by coherent light (under certain condition) all produce squeezed light, 
with a maximum quadrature squeezing of $50\%$ [12--14]. 

The usual commutation relation, $[\hat{a}, \hat{a}^{\dagger}]=1$, has been widely used in the calculation of the quadrature variance or
quadrature squeezing for a cavity mode which is interacting with a two-level atom [15--20]. However, according to the
analysis presented in Ref. [21], the usual commutation relation does not hold for a cavity mode which is interacting with a two-level 
atom. The quadrature variance or the quadrature squeezing calculated employing the usual commutation relation for a cavity mode, which 
is interacting with a two-level atom, cannot therefore be correct.

We seek here to obtain the quadrature squeezing in a cavity mode driven by coherent light and interacting 
with a two-level atom, applying the appropriate commutation relation for the cavity mode operators. We consider the case in which 
the system is coupled to a vacuum reservoir. Moreover, we carry out our analysis by putting the noise operators associated with the 
vacuum reservoir in normal order and without considering the interaction of the two-level atom with the vacuum reservoir outside the 
cavity. Using the steady-state solution of the pertinent quantum Langevin equation, we calculate the quadrature squeezing in the
cavity mode. In addition, we intend to analyze the quantum properties of a pair of superposed cavity modes, each 
driven by coherent light and interacting with a two-level atom. We define the annihilation operator representing the 
superposed cavity modes in terms of the annihilation operators representing the separate cavity modes. We then determine the quantum 
Langevin equation for this operator. Applying the steady-state solution of the resulting equation, we calculate the 
quadrature squeezing.
\section{Operator Dynamics}
We seek to obtain the equations of evolution for the atomic and cavity mode operators with the cavity mode driven by coherent light 
and interacting with a two-level atom. We consider the case in which the cavity mode is coupled to a vacuum reservoir and intend to 
carry out our calculation by putting the noise operators associated with the vacuum reservoir in normal order.
The interaction of the cavity mode with the atom can be described at resonance by the Hamiltonian
% \begin{figure}[bt]
% \begin{center}
% %\includegraphics{}
% \setlength{\unitlength}{0.18mm}
% \begin{tikzpicture}
%    \node (A) at (-1,2){};
%     \node (B) at (1,2) {};
%      \draw [ultra thick, ->]  -| (A) -- (B) node [midway, above] {$\varepsilon$};
%   \end{tikzpicture}
%   \hspace*{-5mm}
% \begin{tikzpicture}
%     	\draw  (1,7)--(1,3);
% \end{tikzpicture}
% \begin{tikzpicture}
%     	\draw [ultra thick] (1,7)--(1,3);
% \end{tikzpicture}
% \begin{picture}(400,250)
% \put(75,30){\line(1,0){130}}
% \put(75,170){\line(1,0){130}}
% %\put(120,200){\vector(0,-1){190}}
% \put(140,170){\vector(0,-1){140}}
% %\put(97,120){$\gamma$}
% \put(142,120){$\gamma_{c}$}
% \put(220,165){$|a\rangle$}
% \put(220,25){$|b\rangle$}
%  \end{picture}
%  \hspace*{-20mm}
%  \begin{tikzpicture}
%     	\draw  (1,7) -- (1,3);
% \end{tikzpicture}
% \begin{tikzpicture}
%     	\draw [ultra thick] (1,7)--(1,3);
% \end{tikzpicture}
% \begin{tikzpicture}
%     	\draw (1,7)--(1,3);
% \end{tikzpicture}
% \hspace*{-5mm}
% \begin{tikzpicture}
%    \node (A) at (-1,2) {};
%     \node (B) at (1,2) {};
%      \draw [ultra thick, ->]  -| (A) -- (B) node [midway, above] {$\kappa$};
%   \end{tikzpicture}
% \end{center}
%\begin{figure}[bt]
%\centering
%\includegraphics[height=9cm,width=12cm]{system}
\begin{figure}
\begin{center}
\begin{picture}(250,180)(-180,-100)
\put(-60,40){\linethickness{0.4mm}\line(1,0){70}}\put(10,38){$|a\rangle$}
%\put(-60,55.5){\line(1,0){60}}
%\put(-3,55){\linethickness{0.4mm}\vector(0,-1){40}}\put(-1,27.5){$\omega_{a}$}
%\put(-40,55){\linethickness{0.4mm}\vector(0,-1){100}}\put(-40,27.5){$\gamma$}
%\put(-40,55){\linethickness{0.4mm}\vector(0,-1){100}}\put(-50,27.5){$\gamma_c$}
%\put(-19,55){\linethickness{0.4mm}\vector(0,-1){40}}\put(-18,27.5){$\gamma_c$}
%\put(-30,55){\linethickness{0.4mm}\vector(0,-1){40}}\put(-30,27.5){$\gamma$}
%\put(-60,15){\linethickness{0.4mm}\line(1,0){70}}\put(10,12.5){$|b\rangle$}
%\put(-19,15){\linethickness{0.4mm}\vector(0,-1){60}}\put(-18,-22.5){$\gamma_c$}
\put(-30,40){\linethickness{0.4mm}\vector(0,-1){85}}\put(-28,-8.5){$\gamma_{c}$}
%\put(-3,15){\linethickness{0.4mm}\vector(0,-1){60}}\put(-1,-22.5){$\omega_{b}$}
\put(-60,-45){\linethickness{0.4mm}\line(1,0){70}}\put(10,-47.5){$|b\rangle$}
%\put(-52,-45){\linethickness{0.4mm}\vector(0,1){100}}
%\put(-60,-45.5){\line(1,0){60}}
\put(-90,-60){\linethickness{1mm}\line(0,1){135}}
\put(-95,-60){\linethickness{0.4mm}\line(0,1){135}}
\put(-150,10){\linethickness{0.4mm}\vector(1,0){55}}\put(-125,14.5){$\bf{\varepsilon}$}
%\put(-120,10){\linethickness{0.4mm}\vector(1,0){40}}\put(-105,14.5){$\bf{\hat b}$}
\put(40,-60){\linethickness{0.58mm}\line(0,1){135}}
\put(44,-60){\linethickness{0.5mm}\line(0,1){135}}
\put(44,10){\linethickness{0.4mm}\vector(1,0){60}}\put(72,14.5){$\bf{\kappa}$}
\end{picture}
\caption{\footnotesize Schematic representation of a two-level atom, a driving coherent light,
and a vacuum reservoir.}
\end{center}
\end{figure}
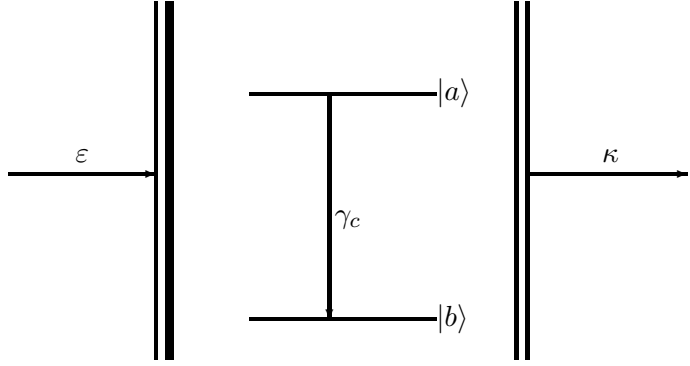 
% \end{figure}\label{fa}
\begin{equation}\label{1}
 \hat{H}'=ig(\hat{\sigma}^{\dag}\hat{a}-\hat{a}^{\dag}\hat{\sigma}),
\end{equation}
where\\
\begin{equation}\label{2}
 \hat{\sigma}=|b\rangle\langle a|,
\end{equation}
is a lowering atomic operator, $\hat{a}$ is the annihilation operator for the cavity mode and $g$ is the
coupling constant between the atom and the cavity mode. On the other hand, the interaction of the cavity mode with the driving coherent 
light can be described by the Hamiltonian
\begin{equation}\label{3}
 \hat{H}''=i\lambda(\hat{b}\hat{a}^{\dagger}-\hat{b}^{\dagger}\hat{a}),
\end{equation}
where $\lambda$ is the coupling constant and $\hat{b}$ is the annihilation operator for the coherent light. 
In order to have a manageable mathematical analysis,  we now replace the operator $\hat{b}$ by a real and
constant c-number $\beta$. Then making use of this replacement, Eq.~\eqref{3} can be put in the form
\begin{equation}\label{4}
 \hat{H}''=i\varepsilon(\hat{a}^{\dagger}-\hat{a}),
\end{equation}
in which
\begin{equation}\label{5}
\varepsilon=\lambda\beta.
\end{equation}
The interaction of the cavity mode with the two-level atom and the driving coherent light can thus be described by the Hamiltonian
\begin{equation}\label{6}
 \hat{H}=ig(\hat{\sigma}^{\dag}\hat{a}-\hat{a}^{\dag}\hat{\sigma})+i\varepsilon(\hat{a}^{\dagger}-\hat{a}).
\end{equation}

Since we carry out our calculation by putting the noise operators associated with the vacuum reservoir in normal order,
the noise operators will not have any effect on the dynamics of the cavity mode operators.
As a result of this, we can drop the noise operator and write the quantum Langevin equation for the operator $\hat{a}$ as
\begin{equation}\label{7}
\frac{d\hat{a}}{dt}=-\frac{\kappa}{2}\hat{a}-i[\hat{a},\hat{H}],
\end{equation}
where $\kappa$ is the cavity damping constant. Now making use of Eq.~\eqref{6}, we get
\begin{equation}\label{8}
\frac{d\hat{a}}{dt}=-\frac{\kappa}{2}\hat{a}-g\hat{\sigma}+\varepsilon.
\end{equation}
Moreover, employing the relation
\begin{equation}\label{9}
\frac{d}{dt}\langle\hat{B}\rangle=-i\langle[\hat{B},\hat{H}]\rangle
\end{equation}
along with ~\eqref{6}, we obtain
\begin{equation}\label{10}
 \frac{d}{dt}\langle\hat{\sigma}\rangle=g\langle(\hat{\eta}_{b}-\hat{\eta}_{a})\hat{a}\rangle,
\end{equation}
\begin{equation}\label{11}
 \frac{d}{dt}\langle\hat{\eta}_{a}\rangle=g\langle\hat{\sigma}^{\dagger}\hat{a}\rangle+g\langle\hat{a}^{\dagger}
 \hat{\sigma}\rangle,
\end{equation}
\begin{equation}\label{12}
 \frac{d}{dt}\langle\hat{\eta}_{b}\rangle=-g\langle\hat{\sigma}^{\dagger}\hat{a}\rangle
 -g\langle\hat{a}^{\dagger}\hat{\sigma}\rangle,
\end{equation}
where
\begin{equation}\label{13}
\hat{\eta}_{a}=|a\rangle\langle a|
\end{equation}
and
\begin{equation}\label{14}
\hat{\eta}_{b}=|b\rangle\langle b|.
\end{equation}

We see that Eqs.~\eqref{10} --~\eqref{12} are coupled nonlinear differential equations and in view of this, 
it is difficult to find the time dependent solutions of these equations. We can avoid this problem by 
applying the large-time approximation scheme $[22]$. Then application of this approximation to Eq.~\eqref{8} yields
\begin{equation}\label{15}
\hat{a}(t)=-\frac{2g}{\kappa}\hat{\sigma}(t)+\frac{2\varepsilon}{\kappa}.
\end{equation}
With the aid of ~\eqref{15} (with the time argument suppressed), the aforementioned equations can be rewritten as
\begin{equation}\label{16}
 \frac{d}{dt}\langle\hat{\sigma}\rangle=-\frac{1}{2}\gamma_{c}\langle\hat{\sigma}\rangle
 +\frac{2g\varepsilon}{\kappa}[\langle\hat{\eta}_{b}\rangle-\langle\hat{\eta}_{a}\rangle],
\end{equation}
\begin{equation}\label{17}
 \frac{d}{dt}\langle\hat{\eta}_{a}\rangle=-\gamma_{c}\langle\hat{\eta}_{a}\rangle
 +\frac{2g\varepsilon}{\kappa}[\langle\hat{\sigma}^{\dagger}\rangle+\langle\hat{\sigma}\rangle],
\end{equation}
 \begin{equation}\label{18}
 \frac{d}{dt}\langle\hat{\eta}_{b}\rangle=\gamma_{c}\langle\hat{\eta}_{a}\rangle
 -\frac{2g\varepsilon}{\kappa}[\langle\hat{\sigma}^{\dagger}\rangle+\langle\hat{\sigma}\rangle],
\end{equation}
where
\begin{equation}\label{19}
\gamma_{c}=\frac{4g^{2}}{\kappa}
\end{equation}
is the stimulated emission decay constant. Moreover, with the aid of the identity 
\begin{equation}\label{20}
\hat{\eta}_{a}+\hat{\eta}_{b}=\hat{I},
\end{equation}
one can write that
\begin{equation}\label{21}
\langle\hat{\eta}_{a}\rangle+\langle\hat{\eta}_{b}\rangle=1.
\end{equation}
Here $\langle\hat{\eta}_{a}\rangle$ and $\langle\hat{\eta}_{b}\rangle$ represent the probabilities
for the atom to be in the upper and lower levels, respectively. 
In addition, using the definition given by Eq.~\eqref{2}, we find that
\begin{equation}\label{22}
\hat{\sigma}^{\dagger}\hat{\sigma}=\hat{\eta}_{a}
\end{equation}
and 
\begin{equation}\label{23}
\hat{\sigma}\hat{\sigma}^{\dagger}=\hat{\eta}_{b}.
\end{equation}
We note that the steady-state solutions of Eqs.~\eqref{16} --~\eqref{18} are given by
\begin{equation}\label{24}
 \langle\hat{\sigma}\rangle=\frac{4g\varepsilon}{\kappa\gamma_{c}}
  [\langle\hat{\eta}_{b}\rangle-\langle\hat{\eta}_{a}\rangle],
\end{equation}
\begin{equation}\label{25}
 \langle\hat{\eta}_{a}\rangle=\frac{2g\varepsilon}{\kappa\gamma_{c}}[\langle\hat{\sigma}\rangle
 +\langle\hat{\sigma}^{\dagger}\rangle],
\end{equation}
and
\begin{equation}\label{26}
\langle\hat{\eta}_{b}\rangle=1-\frac{2g\varepsilon}{\kappa \gamma_{c}}[\langle\hat{\sigma}\rangle
+\langle\hat{\sigma}^{\dagger}\rangle].
\end{equation}
Upon substituting ~\eqref{24} and its complex conjugate into Eqs.~\eqref{25} and ~\eqref{26}, we obtain
 \begin{equation}\label{27}
 \langle\hat{\eta}_{a}\rangle=\frac{4\varepsilon^{2}}{\kappa\gamma_{c}}[\langle\hat{\eta}_{b}\rangle
 -\langle\hat{\eta}_{a}\rangle]
\end{equation}
and
\begin{equation}\label{28}
 \langle\hat{\eta}_{b}\rangle=1-\frac{4\varepsilon^{2}}{\kappa \gamma_{c}}[\langle\hat{\eta}_{b}\rangle
 -\langle\hat{\eta}_{a}\rangle].
\end{equation}
With the aid of the relation given by ~\eqref{21}, Eqs.~\eqref{27} and ~\eqref{28} can be expressed as
\begin{equation}\label{29}
 \langle\hat{\eta}_{a}\rangle=\frac{4\varepsilon^{2}}{8\varepsilon^{2}+\kappa\gamma_{c}}
\end{equation}
and
\begin{equation}\label{30}
 \langle\hat{\eta}_{b}\rangle=\frac{4\varepsilon^{2}+\kappa\gamma_{c}}{8\varepsilon^{2}+\kappa\gamma_{c}}.
\end{equation}
Finally, on introducing Eqs.~\eqref{29} and ~\eqref{30} into ~\eqref{24}, we find
\begin{equation}\label{31}
 \langle\hat{\sigma}\rangle=\frac{4g\varepsilon}{8\varepsilon^{2}+\kappa\gamma_{c}}.
\end{equation}
\section{Quadrature Squeezing}
In this section we seek to calculate the quadrature squeezing of the cavity mode. To this end, we first obtain the variance
for the plus and minus quadratures. From the results we get, we determine the quadrature variance for the vacuum state. 
We then calculate the quadrature squeezing relative to the quadrature variance of the vacuum state. 

The squeezing properties of the cavity mode are described by two quadrature operators defined by
\begin{equation}\label{32}
\begin{split}
\hat{a}_{+}=\hat{a}^{\dagger}+\hat{a}
\end{split}
\end{equation}
and
\begin{equation}\label{33}
\begin{split}
\hat{a}_{-}=i(\hat{a}^{\dagger}-\hat{a}).
\end{split}
\end{equation}
One can easily check that
\begin{equation}\label{34}
\begin{split}
[\hat{a}_{-},\hat{a}_{+}]=-2i[\hat{a},\hat{a}^{\dagger}].
\end{split}
\end{equation}
On the other hand, the steady-state solution of Eq.~\eqref{8} is given by 
\begin{equation}\label{35}
\hat{a}=-\frac{2g}{\kappa}\hat{\sigma}+\frac{2\varepsilon}{\kappa}
\end{equation}
and employing this result, we get
\begin{equation}\label{36}
\begin{split}
[\hat{a},\hat{a}^{\dagger}]=\frac{\gamma_{c}}{\kappa}(\hat{\eta}_{b}-\hat{\eta}_{a}).
\end{split}
\end{equation}
In view of this relation, ~\eqref{34} becomes
\begin{equation}\label{37}
\begin{split}
[\hat{a}_{-},\hat{a}_{+}]=2i\frac{\gamma_{c}}{\kappa}(\hat{\eta}_{a}-\hat{\eta}_{b}).
\end{split}
\end{equation}
Using this result, it can be established that $[14]$
\begin{equation}\label{38}
\begin{split}
\Delta {a}_{+}\Delta {a}_{-}\geq\frac{\gamma_{c}}{\kappa}\Big|\langle\hat{\eta}_{a}\rangle-\langle\hat{\eta}_{b}\rangle\Big|.
\end{split}
\end{equation}
Now applying Eqs.~\eqref{29} and ~\eqref{30}, we find
\begin{equation}\label{39}
\begin{split}
\Delta {a}_{+}\Delta {a}_{-}\geq f_{a}(\varepsilon),
\end{split}
\end{equation}
where
\begin{equation}\label{40}
f_{a}(\varepsilon)=\frac{\gamma_{c}^{2}}{8\varepsilon^{2}+\gamma_{c}\kappa}.
\end{equation}
Here Eq.~\eqref{39} represents the uncertainty relation for the quadrature operators. 
\subsection{Quadrature variance}
The quadrature variance of the cavity mode is expressible as
\begin{equation}\label{41}
\begin{split}
(\Delta a_{\pm})^{2}=\langle\hat{a}^{\dagger}\hat{a}\rangle
+\langle\hat{a}\hat{a}^{\dagger}\rangle\pm\langle\hat{a}^{\dagger2}
\rangle\pm\langle\hat{a}^{2}\rangle\mp\langle\hat{a}^{\dagger}\rangle^{2}
\mp\langle\hat{a}\rangle^{2}-2\langle\hat{a}^{\dagger}\rangle\langle\hat{a}\rangle.
\end{split}
\end{equation}
Applying Eq.~\eqref{35} once more, we get
\begin{equation}\label{42}
\langle\hat{a}^{\dagger}\hat{a}\rangle=\frac{\gamma_{c}}{\kappa}\langle\hat{\sigma}^{\dagger}\hat{\sigma}\rangle
-\frac{4g\varepsilon}{\kappa^{2}}[\langle\hat{\sigma}^{\dagger}\rangle
+\langle\hat{\sigma}\rangle]+\frac{4\varepsilon^{2}}{\kappa^{2}}.
\end{equation}
Upon substituting ~\eqref{31} and its complex conjugate into Eq.~\eqref{42}, we obtain
\begin{equation}\label{43}
\langle\hat{a}^{\dagger}\hat{a}\rangle=\frac{\gamma_{c}}{\kappa}\langle\hat{\sigma}^{\dagger}\hat{\sigma}\rangle
-\frac{\gamma_{c}}{\kappa}[\frac{8\varepsilon^{2}}{8\varepsilon^{2}+\kappa\gamma_{c}}]
+\frac{4\varepsilon^{2}}{\kappa^{2}}.
\end{equation}
With the aid of ~\eqref{22}, Eq.~\eqref{43} can be put in the form 
\begin{equation}\label{44}
\langle\hat{a}^{\dagger}\hat{a}\rangle=\frac{\gamma_{c}}{\kappa}\langle\hat{\eta}_{a}\rangle
-\frac{\gamma_{c}}{\kappa}[\frac{8\varepsilon^{2}}{8\varepsilon^{2}+\kappa\gamma_{c}}]
+\frac{4\varepsilon^{2}}{\kappa^{2}},
\end{equation}
so that in view of ~\eqref{29}, there follows
\begin{equation}\label{45}
\langle\hat{a}^{\dagger}\hat{a}\rangle=\frac{\gamma_{c}}{\kappa}[\frac{4\varepsilon^{2}}{8\varepsilon^{2}+\kappa\gamma_{c}}]
-\frac{\gamma_{c}}{\kappa}[\frac{8\varepsilon^{2}}{8\varepsilon^{2}+\kappa\gamma_{c}}]
+\frac{4\varepsilon^{2}}{\kappa^{2}}
\end{equation}
or
\begin{equation}\label{46}
\bar{n}=\frac{4\varepsilon^{2}}{\kappa^{2}}-\frac{\gamma_{c}}{\kappa}[\frac{4\varepsilon^{2}}{8\varepsilon^{2}
+\kappa\gamma_{c}}].
\end{equation}
This is the steady-state mean photon number of the cavity mode with the cavity mode driven by coherent light and interacting with 
the two-level atom. We notice from Eq.~\eqref{45} that the first and second terms represent, respectively, the mean number of photons 
emitted and absorbed and the third term describes the mean number of photons due to the driving coherent light. Moreover, we see 
from ~\eqref{46} that the mean number of photons emitted is less than the mean number of photons absorbed. This must be due to the fact 
that the probability for the atom to be in the lower level is greater than that in the upper level. 

Moreover, applying Eq.~\eqref{35} once again, we obtain
\begin{equation}\label{47}
\begin{split}
\langle\hat{a}\hat{a}^{\dagger}\rangle=\frac{\gamma_{c}}{\kappa}\langle\hat{\sigma}\hat{\sigma}^{\dagger}\rangle
-\frac{4g\varepsilon}{\kappa^{2}}[\langle\hat{\sigma}\rangle+\langle\hat{\sigma}^{\dagger}\rangle]+\frac{4\varepsilon^{2}}{\kappa^{2}}.
\end{split}
\end{equation}
Then on account of Eq.~\eqref{30}, ~\eqref{47} takes the form
\begin{equation}\label{48}
\begin{split}
\langle\hat{a}\hat{a}^{\dagger}\rangle=\frac{\gamma_{c}}{\kappa}\langle\hat{\eta}_{b}\rangle
-\frac{4g\varepsilon}{\kappa^{2}}[\langle\hat{\sigma}\rangle+\langle\hat{\sigma}^{\dagger}\rangle]
+\frac{4\varepsilon^{2}}{\kappa^{2}}.
\end{split}
\end{equation}
On introducing ~\eqref{31} and its complex conjugate into this equation, we find
\begin{equation}\label{49}
\begin{split}
\langle\hat{a}\hat{a}^{\dagger}\rangle=\frac{\gamma_{c}}{\kappa}\langle\hat{\eta}_{b}\rangle
-\frac{\gamma_{c}}{\kappa}[\frac{8\varepsilon^{2}}{8\varepsilon^{2}+\kappa\gamma_{c}}]
+\frac{4\varepsilon^{2}}{\kappa^{2}}.
\end{split}
\end{equation}
Now with the aid of ~\eqref{30}, Eq.~\eqref{49} can be put in the form 
\begin{equation}\label{50}
\begin{split}
\langle\hat{a}\hat{a}^{\dagger}\rangle=\frac{\gamma_{c}}{\kappa}
[\frac{\kappa\gamma_{c}-4\varepsilon^{2}}{8\varepsilon^{2}+\kappa\gamma_{c}}]
+\frac{4\varepsilon^{2}}{\kappa^{2}}.
\end{split}
\end{equation}
Furthermore, using ~\eqref{35} once more, we see that
\begin{equation}\label{51}
\begin{split}
\langle\hat{a}^{2}\rangle=\frac{\gamma_{c}}{\kappa}\langle\hat{\sigma}^{2}\rangle
-\frac{8g\varepsilon}{\kappa^{2}}\langle\hat{\sigma}\rangle+\frac{4\varepsilon^{2}}{\kappa^{2}},
\end{split}
\end{equation}
so that in view of the fact that $\langle\hat{\sigma}^{2}\rangle=0$, there follows
\begin{equation}\label{52}
\begin{split}
\langle\hat{a}^{2}\rangle=-\frac{8g\varepsilon}{\kappa^{2}}\langle\hat{\sigma}\rangle+\frac{4\varepsilon^{2}}{\kappa^{2}}.
\end{split}
\end{equation}
Now upon substituting Eq.~\eqref{31} into ~\eqref{52}, we have
\begin{equation}\label{53}
\begin{split}
\langle\hat{a}^{2}\rangle=\frac{4\varepsilon^{2}}{\kappa^{2}}-\frac{\gamma_{c}}{\kappa}
[\frac{8\varepsilon^{2}}{8\varepsilon^{2}+\kappa\gamma_{c}}].
\end{split}
\end{equation}
Moreover, applying the expectation value of Eq.~\eqref{35}, we obtain
\begin{equation}\label{54}
\begin{split}
\langle\hat{a}\rangle^{2}=\frac{\gamma_{c}}{\kappa}\langle\hat{\sigma}\rangle^{2}-
\frac{8g\varepsilon}{\kappa^{2}}\langle\hat{\sigma}\rangle+\frac{4\varepsilon^{2}}{\kappa^{2}}
\end{split}
\end{equation}
and on account of ~\eqref{31}, we get
\begin{equation}\label{55}
\begin{split}
\langle\hat{a}\rangle^{2}&=\frac{4\gamma_{c}^{2}\varepsilon^{2}}{[8\varepsilon^{2}+\kappa\gamma_{c}]^{2}}
-\frac{\gamma_{c}}{\kappa}[\frac{8\varepsilon^{2}}{8\varepsilon^{2}+\kappa\gamma_{c}}]
+\frac{4\varepsilon^{2}}{\kappa^{2}}.
\end{split}
\end{equation}
Finally, on combining Eqs.~\eqref{45}, ~\eqref{50}, ~\eqref{53}, and ~\eqref{55}, we arrive at 
\begin{equation}\label{56}
\begin{split}
(\Delta a_{+})^{2}&=\frac{\gamma_{c}}{\kappa}-\frac{16\gamma_{c}^{2}\varepsilon^{2}}{[8\varepsilon^{2}
+\kappa\gamma_{c}]^{2}}
\end{split}
\end{equation}
and
\begin{equation}\label{57}
\begin{split}
(\Delta a_{-})^{2}=\frac{\gamma_{c}}{\kappa}.
\end{split}
\end{equation}
Upon setting $\varepsilon=0$ in Eq.~\eqref{56}, we get
\begin{equation}\label{58}
\begin{split}
(\Delta a_{\pm})_{\nu}^{2}=\frac{\gamma_{c}}{\kappa}.
 \end{split}
\end{equation}
This represents the quadrature variance for a cavity mode in a vacuum state. Comparison of Eq.~\eqref{56} with
~\eqref{58} shows that the variance for the plus quadrature is less than the vacuum-state quadrature variance. 
This indicates that the cavity mode is in a squeezed state. Moreover, using Eqs.~\eqref{56} and ~\eqref{57}, we see that
\begin{equation}\label{59}
\Delta a_{+}=\sqrt{\frac{\gamma_{c}}{\kappa}-\frac{16\gamma_{c}^{2}\varepsilon^{2}}{[8\varepsilon^{2}
+\kappa\gamma_{c}]^{2}}}
\end{equation}
and
\begin{equation}\label{60}
\Delta a_{-}=\sqrt{\frac{\gamma_{c}}{\kappa}}.
\end{equation}
Thus the product of ~\eqref{59} and ~\eqref{60} can be written as
\begin{equation}\label{61}
\Delta a_{+}\Delta a_{-}=f_{b}(\varepsilon),
\end{equation}
where 
\begin{equation}\label{62}
f_{b}(\varepsilon)=\sqrt{\frac{\gamma_{c}^{2}}{\kappa^{2}}-\frac{16\gamma_{c}^{3}\varepsilon^{2}}{\kappa
[8\varepsilon^{2}+\kappa\gamma_{c}]^{2}}}.
\end{equation}
\begin{figure}[bt]
\centering
{\includegraphics[height=9cm,width=12cm]{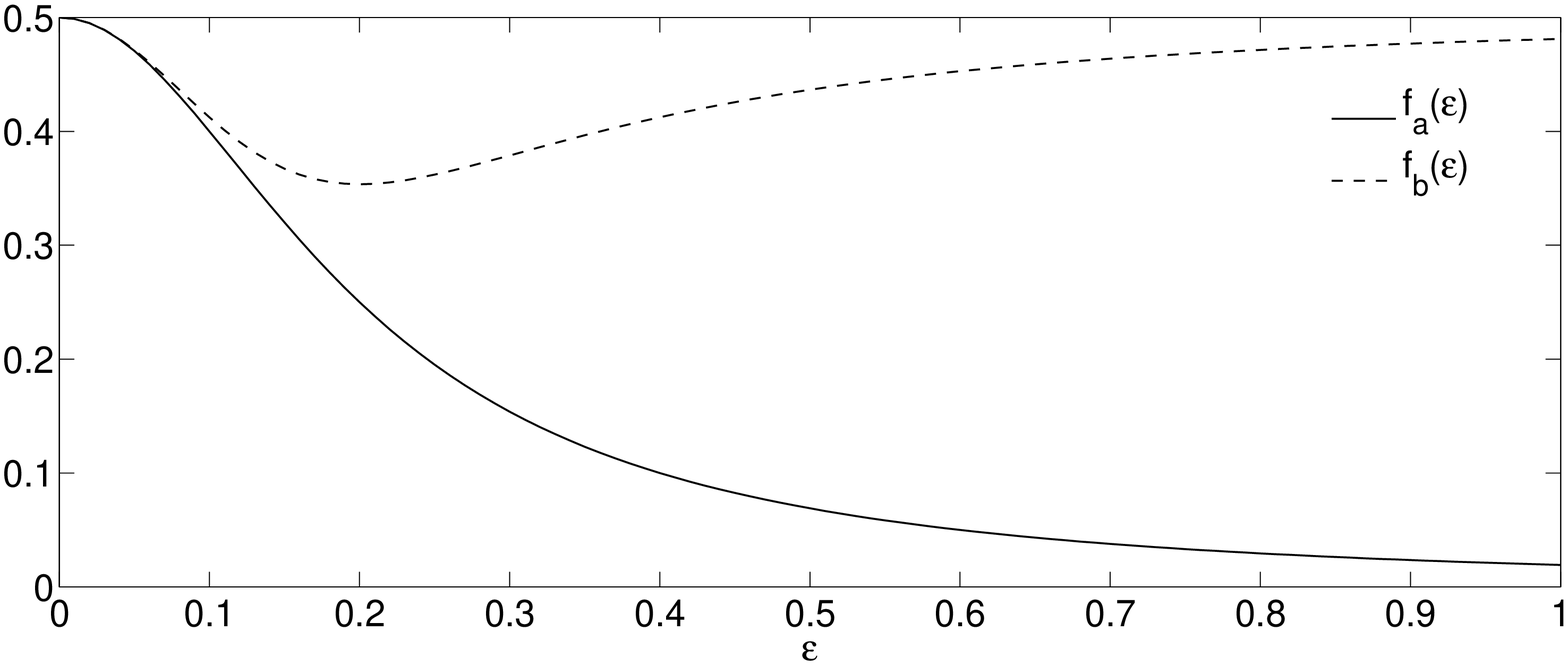}}
\caption{\footnotesize{The plots of Eqs.~\eqref{40} (solid curve) and  ~\eqref{62} (dashed curve) versus $\varepsilon$ 
for $\gamma_{c}=0.4$ and $\kappa=0.8$}.}
\label{f00}
\end{figure}
We see from the plots in Fig.~\ref{f00} that $f_{b}(\varepsilon)=f_{a}(\varepsilon)$ in the interval $0<\varepsilon<0.07$.
Moreover, for $\varepsilon>0.07$, $f_{b}(\varepsilon)>f_{a}(\varepsilon)$. From this, we see that the uncertainty relation for
the quadrature operators is satisfied.
\subsection{Quadrature squeezing}
We define the quadrature squeezing relative to the quadrature variance of the vacuum state by $[14]$
\begin{equation}\label{63}
\begin{split}
S=\frac{(\Delta a_{+})_{\nu}^{2}-(\Delta a_{+})^{2}}{(\Delta a_{+})_{\nu}^{2}}.
\end{split}
\end{equation}
Hence in view of ~\eqref{56} and ~\eqref{58}, this equation takes the form
\begin{equation}\label{64}
\begin{split}
S=\frac{16\gamma_{c}\kappa\varepsilon^{2}}{[8\varepsilon^{2}+\kappa\gamma_{c}]^{2}}.
\end{split}
\end{equation}
\begin{figure}[bt]
\centering
\includegraphics[height=9cm,width=12cm]{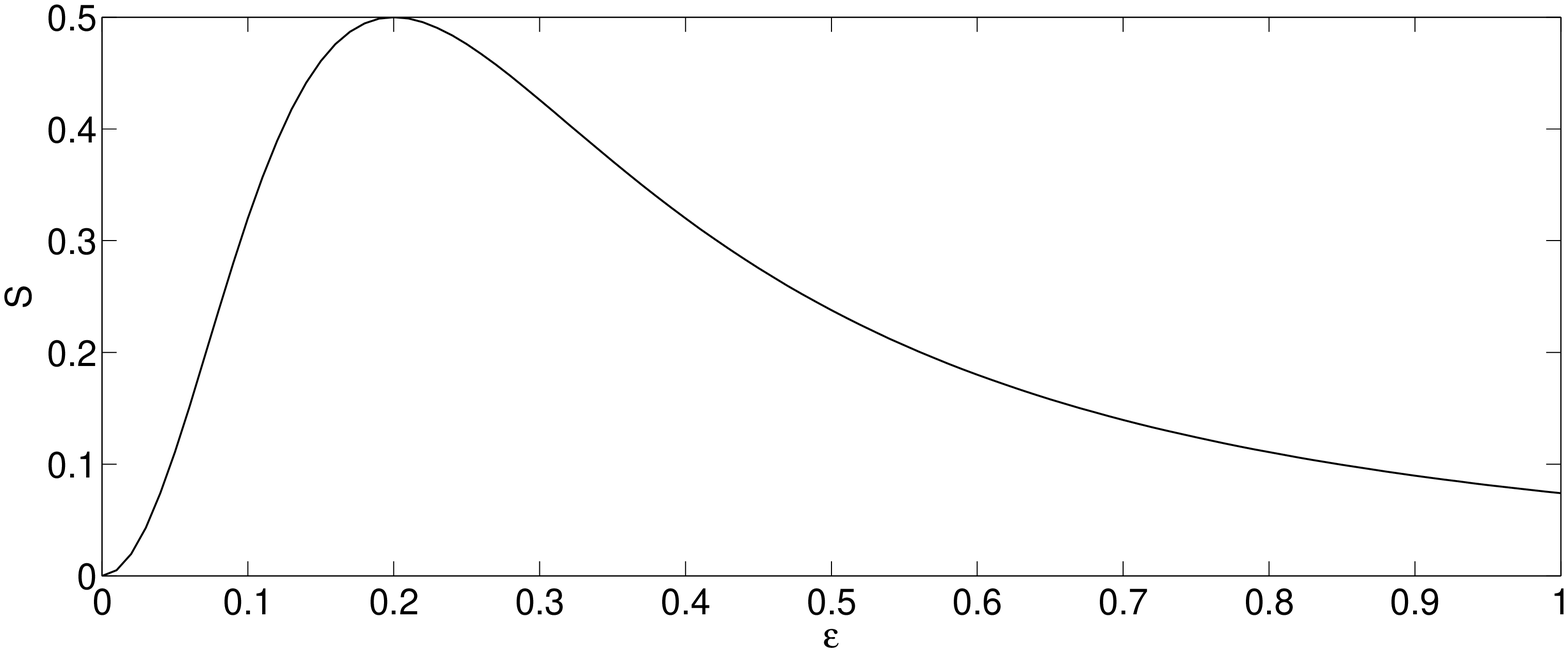}
\caption{\footnotesize{A plot of Eq.~\eqref{64} versus $\varepsilon$ for $\gamma_{c}=0.4$ and $\kappa=0.8$}.}
\label{f0}
\end{figure}
This represents the quadrature squeezing for the cavity mode. The plot in Fig.~\ref{f0} shows that the quadrature squeezing 
increases with the amplitude of the driving coherent
light until it reaches a maximum value of $50\%$ below the vacuum-state level.
\section{Superposed Cavity Modes}
In this section we intend to analyze the quantum properties of a pair of superposed cavity modes, each 
driven by coherent light and interacting with a two-level atom. We wish to represent the two cavity modes by 
the annihilation operators $\hat{a}$ and $\hat{b}$. Using these operators, we define the annihilation operator representing the 
superposed cavity modes. We then determine the quantum Langevin equation for the superposed cavity modes. Applying the steady-state 
solution of the resulting equation, we calculate the quadrature squeezing.

According to Fesseha $[21]$, we can define the annihilation operator representing the superposition of light modes $a$ and $b$ by
\begin{equation}\label{65}
 \hat{c}=\hat{a}+i\hat{b}.
\end{equation}
Then the equation of evolution for $\hat{c}$ becomes
\begin{equation}\label{66}
 \frac{d\hat{c}}{dt}=\frac{d\hat{a}}{dt}+i\frac{d\hat{b}}{dt}.
\end{equation}
According to Eq.~\eqref{8}, one can write the equations of evolution for the cavity modes as
\begin{equation}\label{67}
 \frac{d\hat{a}}{dt}=-\frac{\kappa}{2}\hat{a}-g\hat{\sigma}_{a}+\varepsilon
\end{equation}
and
\begin{equation}\label{68}
 \frac{d\hat{b}}{dt}=-\frac{\kappa}{2}\hat{b}-g\hat{\sigma}_{b}+\varepsilon.
\end{equation}
Now upon substituting ~\eqref{67} and ~\eqref{68} into Eq.~\eqref{66}, we get 
\begin{equation}\label{69}
 \frac{d\hat{c}}{dt}=-\frac{\kappa}{2}[\hat{a}+i\hat{b}]-g[\hat{\sigma}_{a}+i\hat{\sigma}_{b}]+[1+i]\varepsilon.
\end{equation}
In view of ~\eqref{65}, Eq.~\eqref{69} can be rewritten in the form
\begin{equation}\label{70}
 \frac{d\hat{c}}{dt}=-\frac{\kappa}{2}\hat{c}-g\hat{m}+[1+i]\varepsilon,
\end{equation}
with
\begin{equation}\label{71}
 \hat{m}=\hat{\sigma}_{a}+i\hat{\sigma}_{b}.
\end{equation}
Now applying ~\eqref{71}, one can easily establish that
\begin{equation}\label{72}
 \hat{m}^{\dagger}\hat{m}=\hat{\eta}_{a}+\hat{\eta}_{a}'
\end{equation}
and
\begin{equation}\label{73}
 \hat{m}\hat{m}^{\dagger}=\hat{\eta}_{b}+\hat{\eta}_{b}',
\end{equation}
with $\hat{\eta}_{a}$ and $\hat{\eta}_{a}'$ representing the probability for each atom to be in the upper level, and
$\hat{\eta}_{b}$ and $\hat{\eta}_{b}'$ representing the probability for each atom to be in the lower level. 
\subsection{Quadrature variance}
The squeezing properties of the superposed cavity modes are described by two quadrature operators defined by
\begin{equation}\label{74}
 \hat{c}_{+}=\hat{c}^{\dagger}+\hat{c}
\end{equation}
and
\begin{equation}\label{75}
 \hat{c}_{-}=i(\hat{c}^{\dagger}-\hat{c}).
\end{equation}
Using Eqs.~\eqref{74} and ~\eqref{75}, we get
\begin{equation}\label{76}
 [\hat{c}_{-},\hat{c}_{+}]=-2i[\hat{c},\hat{c}^{\dagger}].
\end{equation}
On the other hand, the steady-state solution of ~\eqref{70} is expressible as
\begin{equation}\label{77}
 \hat{c}=-\frac{2g}{\kappa}\hat{m}+\frac{2\varepsilon}{\kappa}[1+i]
\end{equation}
and employing this result, we obtain
\begin{equation}\label{78}
 [\hat{c},\hat{c}^{\dagger}]=\frac{\gamma_{c}}{\kappa}[\hat{m}\hat{m}^{\dagger}-\hat{m}^{\dagger}\hat{m}].
\end{equation}
Now with the aid of the relations given by Eqs.~\eqref{72} and ~\eqref{73}, one easily obtains
\begin{equation}\label{79}
[\hat{c},\hat{c}^{\dagger}]=\frac{\gamma_{c}}{\kappa}[(\hat{\eta}_{b}-\hat{\eta}_{a})+(\hat{\eta}_{b}'-\hat{\eta}_{a}')],
\end{equation}
so that Eq.~\eqref{76} takes the form
\begin{equation}\label{80}
 [\hat{c}_{-},\hat{c}_{+}]=2i\frac{\gamma_{c}}{\kappa}[(\hat{\eta}_{a}-\hat{\eta}_{b})+(\hat{\eta}_{a}'-\hat{\eta}_{b}')].
\end{equation}
On account of this result, we see that 
\begin{equation}\label{81}
 \Delta c_{-}\Delta c_{+}\geq \frac{\gamma_{c}}{\kappa}\Big|\langle\hat{\eta}_{a}\rangle-\langle\hat{\eta}_{b}\rangle
 +\langle\hat{\eta}_{a}'\rangle-\langle\hat{\eta}_{b}'\rangle\Big|.
\end{equation}
According to Eqs.~\eqref{29} and ~\eqref{30}, we can establish that
\begin{equation}\label{82}
 \langle\hat{\eta}_{a}'\rangle=\frac{4\varepsilon^{2}}{8\varepsilon^{2}+\kappa\gamma_{c}},
\end{equation}
and
\begin{equation}\label{83}
 \langle\hat{\eta}_{b}'\rangle=\frac{4\varepsilon^{2}+\kappa\gamma_{c}}{8\varepsilon^{2}
 +\kappa\gamma_{c}},
\end{equation}
with $\langle\hat{\eta}_{a}'\rangle$ and $\langle\hat{\eta}_{b}'\rangle$ 
representing the probabilities for the atom interacting with cavity mode $b$ to be in the upper and lower levels.
Now employing Eqs.~\eqref{29}, ~\eqref{30}, ~\eqref{82}, and ~\eqref{83} in ~\eqref{81}, we arrive at
\begin{equation}\label{84}
 \Delta c_{-}\Delta c_{+}\geq f_{c}(\varepsilon),
\end{equation}
where
\begin{equation}\label{85}
f_{c}(\varepsilon)=\frac{2\gamma_{c}^{2}}{8\varepsilon^{2}+\kappa\gamma_{c}}.
\end{equation}
We thus see that Eq.~\eqref{84} represents the uncertainty relation for the quadrature operators.

The quadrature variance of the superposed cavity modes is expressible as
\begin{equation}\label{86}
 (\Delta c_{\pm})^{2}=\langle\hat{c}^{\dagger}\hat{c}\rangle+\langle\hat{c}\hat{c}^{\dagger}\rangle\pm\langle\hat{c}^{\dagger2}\rangle
 \pm\langle\hat{c}^{2}\rangle\mp\langle\hat{c}\rangle^{\dagger2}\mp\langle\hat{c}\rangle^{2}-2\langle\hat{c}^{\dagger}\rangle
 \langle\hat{c}\rangle.
\end{equation}
Using Eq.~\eqref{77}, we get
\begin{equation}\label{87}
\begin{split}
 \langle\hat{c}^{\dagger}\hat{c}\rangle&=\frac{\gamma_{c}}{\kappa}\langle\hat{m}^{\dagger}\hat{m}\rangle
 -\frac{4g\varepsilon}{\kappa^{2}}[(1+i)\langle\hat{m}^{\dagger}\rangle
 +(1-i)\langle\hat{m}\rangle]+\frac{8\varepsilon^{2}}{\kappa^{2}}
 \end{split}
\end{equation}
aand with the aid of the relation given by Eq.~\eqref{72}, we have
\begin{equation}\label{88}
\begin{split}
 \langle\hat{c}^{\dagger}\hat{c}\rangle&=\frac{\gamma_{c}}{\kappa}[\langle\hat{\eta}_{a}\rangle+\langle\hat{\eta}_{a}'\rangle]
 -\frac{4g\varepsilon}{\kappa^{2}}[(1+i)\langle\hat{m}^{\dagger}\rangle
 +(1-i)\langle\hat{m}\rangle]+\frac{8\varepsilon^{2}}{\kappa^{2}}.
 \end{split}
\end{equation}
Moreover, on taking the expectation value of Eq.~\eqref{71}, we see that
\begin{equation}\label{89}
 \langle\hat{m}\rangle=\langle\hat{\sigma}_{a}\rangle+i\langle\hat{\sigma}_{b}\rangle.
\end{equation}
Upon substituting ~\eqref{89} and its complex conjugate into Eq.~\eqref{88}, we obtain
\begin{equation}\label{90}
\begin{split}
 \langle\hat{c}^{\dagger}\hat{c}\rangle&=\frac{\gamma_{c}}{\kappa}[\langle\hat{\eta}_{a}\rangle+\langle\hat{\eta}_{a}'\rangle]
 -\frac{4g\varepsilon}{\kappa^{2}}[\langle\hat{\sigma}_{a}\rangle
 +\langle\hat{\sigma}_{a}^{\dagger}\rangle+\langle\hat{\sigma}_{b}\rangle
 +\langle\hat{\sigma}_{b}^{\dagger}\rangle\\&+i(\langle\hat{\sigma}_{a}^{\dagger}\rangle-\langle\hat{\sigma}_{a}\rangle)
 +i(\langle\hat{\sigma}_{b}\rangle-\langle\hat{\sigma}_{b}^{\dagger}\rangle)]
 +\frac{8\varepsilon^{2}}{\kappa^{2}}
 \end{split}
\end{equation}
and using the fact that $\langle\hat{\sigma}_{a}\rangle=\langle\hat{\sigma}_{a}^{\dagger}\rangle$ and 
$\langle\hat{\sigma}_{b}\rangle=\langle\hat{\sigma}_{b}^{\dagger}\rangle$, we have
\begin{equation}\label{91}
 \langle\hat{c}^{\dagger}\hat{c}\rangle=\frac{\gamma_{c}}{\kappa}[\langle\hat{\eta}_{a}\rangle+\langle\hat{\eta}_{a}'\rangle]
 -\frac{8g\varepsilon}{\kappa^{2}}[\langle\hat{\sigma}_{a}\rangle+\langle\hat{\sigma}_{b}\rangle]+\frac{8\varepsilon^{2}}{\kappa^{2}}.
\end{equation}
Furthermore, in view of ~\eqref{31}, one can write
\begin{equation}\label{92}
 \langle\hat{\sigma}_{a}\rangle=\frac{4g\varepsilon}{8\varepsilon^{2}+\kappa\gamma_{c}}.
\end{equation}
and
\begin{equation}\label{93}
 \langle\hat{\sigma}_{b}\rangle=\frac{4g\varepsilon}{8\varepsilon^{2}+\kappa\gamma_{c}}.
\end{equation}
Finally, on introducing Eqs.~\eqref{29}, ~\eqref{82}, ~\eqref{92}, and ~\eqref{93} into ~\eqref{91}, we arrive at 
\begin{equation}\label{94}
 \bar{n}_{sup}=2\bar{n},
\end{equation}
with $\bar{n}$ representing the mean photon number for cavity mode $a$ or $b$. We see from Eq.~\eqref{94} that the mean photon number 
of the superposed cavity modes is twice the mean photon number of either of the cavity modes. 

Moreover, appplying ~\eqref{77} once more, we get 
\begin{equation}\label{95}
\begin{split}
 \langle\hat{c}\hat{c}^{\dagger}\rangle&=\frac{\gamma_{c}}{\kappa}\langle\hat{m}\hat{m}^{\dagger}\rangle
 -\frac{4g\varepsilon}{\kappa^{2}}[(1-i)\langle\hat{m}\rangle
 +(1+i)\langle\hat{m}^{\dagger}\rangle]+\frac{8\varepsilon^{2}}{\kappa^{2}}.
 \end{split}
\end{equation}
Now with the aid of the relation given by ~\eqref{73}, we obtain
\begin{equation}\label{96}
\begin{split}
 \langle\hat{c}\hat{c}^{\dagger}\rangle&=\frac{\gamma_{c}}{\kappa}[\langle\hat{\eta}_{b}\rangle+\langle\hat{\eta}_{b}'\rangle]
 -\frac{4g\varepsilon}{\kappa^{2}}[(1-i)\langle\hat{m}\rangle
 +(1+i)\langle\hat{m}^{\dagger}\rangle]+\frac{8\varepsilon^{2}}{\kappa^{2}}.
 \end{split}
\end{equation}
Upon substituting ~\eqref{71} and its complex conjugate into Eq.~\eqref{96} and using the fact that
$\langle\hat{\sigma}_{a}\rangle=\langle\hat{\sigma}_{a}^{\dagger}\rangle$ and $\langle\hat{\sigma}_{b}\rangle
=\langle\hat{\sigma}_{b}^{\dagger}\rangle$, we obtain
\begin{equation}\label{97}
 \langle\hat{c}\hat{c}^{\dagger}\rangle=\frac{\gamma_{c}}{\kappa}[\langle\hat{\eta}_{b}\rangle+\langle\hat{\eta}_{b}'\rangle]
 -\frac{8g\varepsilon}{\kappa^{2}}[\langle\hat{\sigma}_{a}\rangle+\langle\hat{\sigma}_{b}\rangle]
 +\frac{8\varepsilon^{2}}{\kappa^{2}}.
\end{equation}
Now in view of Eqs.~\eqref{30}, ~\eqref{83}, ~\eqref{92}, and ~\eqref{93}, we arrive at
\begin{equation}\label{98}
\begin{split}
 \langle\hat{c}\hat{c}^{\dagger}\rangle&=\frac{8\gamma_{c}\varepsilon^{2}
 +2\kappa\gamma_{c}^{2}}{\kappa[8\varepsilon^{2}+\kappa\gamma_{c}]}
 -\frac{16\gamma_{c}\varepsilon^{2}}{\kappa[8\varepsilon^{2}+\kappa\gamma_{c}]}
 +\frac{8\varepsilon^{2}}{\kappa^{2}}.
 \end{split}
\end{equation}
Moreover, applying Eq.~\eqref{77} once again, we have
\begin{equation}\label{99}
 \langle\hat{c}^{2}\rangle=\frac{\gamma_{c}}{\kappa}\langle\hat{m}^{2}\rangle-\frac{8g\varepsilon}
 {\kappa^{2}}[1+i]\langle\hat{m}\rangle+\frac{4\varepsilon^{2}}{\kappa^{2}}[1+i]^{2}
\end{equation}
and using the fact that $\langle\hat{m}^{2}\rangle=0$, we find
\begin{equation}\label{100}
 \langle\hat{c}^{2}\rangle=\frac{4\varepsilon^{2}}{\kappa^{2}}[1+i]^{2}-\frac{8g\varepsilon}
 {\kappa^{2}}[1+i]\langle\hat{m}\rangle.
\end{equation}
Hence on substituting ~\eqref{71} into Eq.~\eqref{100}, we get
\begin{equation}\label{101}
 \langle\hat{c}^{2}\rangle=\frac{4\varepsilon^{2}}{\kappa^{2}}
 [1+i]^{2}-\frac{8g\varepsilon}
 {\kappa^{2}}[1+i][\langle\hat{\sigma}_{a}\rangle
 +i\langle\hat{\sigma}_{b}\rangle].
\end{equation}
Furthermore, in view of Eqs.~\eqref{92} and ~\eqref{93}, we have
\begin{equation}\label{102}
 \langle\hat{c}^{2}\rangle=\frac{8\varepsilon^{2}i}{\kappa^{2}}-\frac{16\gamma_{c}\varepsilon^{2}i}
 {\kappa[8\varepsilon^{2}+\kappa\gamma_{c}]}.
\end{equation}
On the other hand, upon taking the expectation value of Eq.~\eqref{77} along with ~\eqref{89}, we find
\begin{equation}\label{103}
 \langle\hat{c}\rangle=\frac{2\varepsilon}{\kappa}[1+i]-\frac{2g}{\kappa}[\langle\hat{\sigma}_{a}\rangle
 +i\langle\hat{\sigma}_{b}\rangle]
\end{equation}
 and on substituting ~\eqref{92} and ~\eqref{93} into Eq.~\eqref{103}, one readily obtains
\begin{equation}\label{104}
 \langle\hat{c}\rangle=[\frac{2\varepsilon}{\kappa}-\frac{2\gamma_{c}\varepsilon}
 {8\varepsilon^{2}+\kappa\gamma_{c}}][1+i].
\end{equation}
Now employing Eq.~\eqref{104}, we get 
\begin{equation}\label{105}
 \langle\hat{c}\rangle^{2}=\frac{8\varepsilon^{2}i}{\kappa^{2}}-\frac{16\gamma_{c}\varepsilon^{2}i}
 {\kappa[8\varepsilon^{2}+\kappa\gamma_{c}]}+\frac{8\gamma_{c}^{2}\varepsilon^{2}i}
 {[8\varepsilon^{2}+\kappa\gamma_{c}]^{2}}.
\end{equation}
Finally, in view of Eqs.~\eqref{94}, ~\eqref{98}, ~\eqref{102}, ~\eqref{104}, and ~\eqref{105}, we arrive at
\begin{equation}\label{106}
 (\Delta c_{+})^{2}=\frac{2\gamma_{c}}{\kappa}-\frac{16\gamma_{c}^{2}\varepsilon^{2}}
 {[8\varepsilon^{2}+\kappa\gamma_{c}]^{2}}
\end{equation}
and
\begin{equation}\label{107}
 (\Delta c_{-})^{2}=\frac{2\gamma_{c}}{\kappa}-\frac{16\gamma_{c}^{2}\varepsilon^{2}}
 {[8\varepsilon^{2}+\kappa\gamma_{c}]^{2}}.
\end{equation}

On setting $\varepsilon=0$ in Eqs.~\eqref{106} and ~\eqref{107}, we get the quadrature variance for 
a vacuum state in the form 
\begin{equation}\label{108}
 (\Delta c_{\pm})_{\nu}^{2}=\frac{2\gamma_{c}}{\kappa}.
\end{equation}
Moreover, using Eqs.~\eqref{106} and ~\eqref{107} once more, we get
\begin{equation}\label{109}
 \Delta c_{+}=\sqrt{\frac{2\gamma_{c}}{\kappa}-\frac{16\gamma_{c}^{2}\varepsilon^{2}}
 {[8\varepsilon^{2}+\kappa\gamma_{c}]^{2}}}
\end{equation}
and
\begin{equation}\label{110}
 \Delta c_{-}=\sqrt{\frac{2\gamma_{c}}{\kappa}-\frac{16\gamma_{c}^{2}\varepsilon^{2}}
 {[8\varepsilon^{2}+\kappa\gamma_{c}]^{2}}}.
\end{equation}
Now the product of Eqs.~\eqref{109} and ~\eqref{110} can be written as
\begin{equation}\label{111}
 \Delta c_{-} \Delta c_{+}=f_{d}(\varepsilon),
\end{equation}
with
\begin{equation}\label{112}
 f_{d}(\varepsilon)=\sqrt{\frac{4\gamma_{c}^{2}}{\kappa^{2}}-\frac{64\gamma_{c}^{3}\varepsilon^{2}}
 {\kappa[8\varepsilon^{2}+\kappa\gamma_{c}]^{2}}
 +\frac{256\gamma_{c}^{4}\varepsilon^{4}}{[8\varepsilon^{2}
 +\kappa\gamma_{c}]^{4}}}.
\end{equation}
\begin{figure}[htb]
\begin{center}
 \includegraphics[height=9cm, width=12cm]{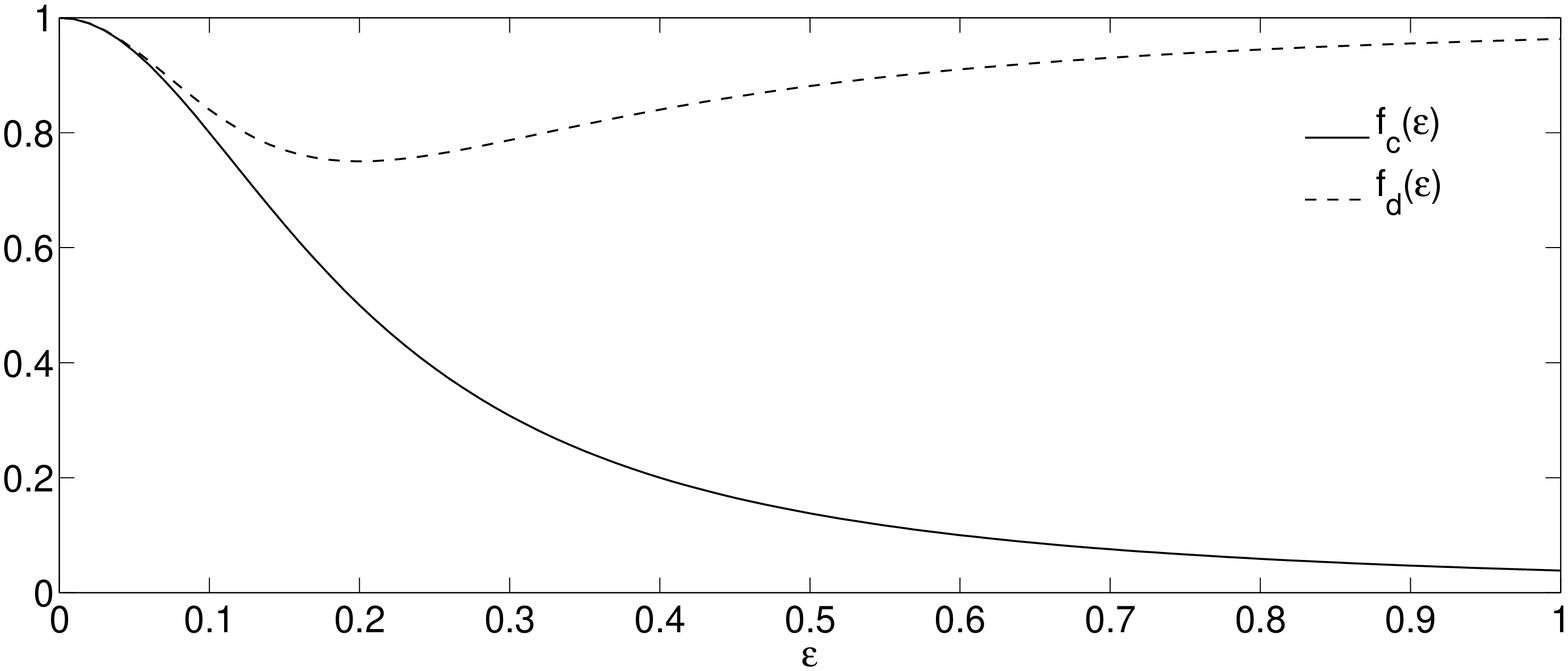}
 \caption{\footnotesize{The plots of Eqs.~\eqref{85} (solid curve) and ~\eqref{112} (dashed curve) versus $\varepsilon$ 
 for $\gamma_{c}=0.4$ and $\kappa=0.8$.}}\label{fj}
\end{center}
\end{figure}
Comparision of ~\eqref{106} and ~\eqref{107} with Eq.~\eqref{108} shows that the superposed cavity modes are in a squeezed
state and the squeezing occurs in both quadratures. Moreover, we note from the plots in Fig.~\ref{fj} that 
$f_{d}(\varepsilon)=f_{c}(\varepsilon)$ in the interval $0<\varepsilon<0.05$. For $\varepsilon>0.05$, 
$f_{d}(\varepsilon)>f_{c}(\varepsilon)$. Based on these results, we observe that the uncertainty relation for the quadrature
operators is satisfied. 
\subsection{Quadrature squeezing}
The quadrature squeezing for the superposed cavity modes is defined relative to the quadrature variance of the vacuum state as
\begin{equation}\label{113}
 S_{sup}=\frac{(\Delta c_{\pm})_{\nu}^{2}-(\Delta c_{\pm})^{2}}{(\Delta c_{\pm})_{\nu}^{2}}.
\end{equation}
Then the squeezing in the plus quadrature is given by
\begin{equation}\label{114}
 S_{+}=\frac{(\Delta c_{+})_{\nu}^{2}-(\Delta c_{+})^{2}}{(\Delta c_{+})_{\nu}^{2}},
\end{equation}
so that in view of ~\eqref{106} and ~\eqref{108}, there follows
\begin{equation}\label{115}
 S_{+}=\frac{S}{2},
\end{equation}
with
\begin{equation}\label{116}
 S=\frac{16\kappa\gamma_{c}\varepsilon^{2}}{[8\varepsilon^{2}+\kappa\gamma_{c}]^{2}}.
\end{equation}
This is the quadrature squeezing for cavity mode $a$ or $b$. Similarly, one easily finds
\begin{equation}\label{117}
 S_{-}=\frac{S}{2}.
\end{equation}
We then see that the sum of the squeezing in the plus and minus quadratures to be
\begin{equation}\label{136}
 S_{+}+S_{-}=S.
\end{equation}
% \begin{figure}[htb]
%  \begin{center}
%  \includegraphics[height=10cm,width=14cm]{Squeezingsa}
%  \caption{\small {The plot of Eq.~\eqref{133} versus $\varepsilon$ for $\gamma_{c}=0.4$ and $\kappa=0.8$.}}\label{fk}
% \end{center}
% \end{figure}
We note from Eq.~\eqref{136} that the sum of the squeezing in the plus and minus quadratures is the same as the squeezing for 
light mode $a$ or $b$. 
\section{Conclusion}
We have seen that the mean number of the emitted photons is less than that of the absorbed photons. 
This must be due to the fact that the probability for the atom to be in the lower level is greater than that in the upper level.
On the other hand, our analysis shows that the cavity mode is in a squeezed state and the squeezing occurs in the plus quadrature. 
Moreover, we have seen that the quadrature squeezing depends on the amplitude of the driving coherent light. In addition, we have 
observed that the quadrature squeezing increases with the amplitude of 
the driving coherent light until it reaches a maximum value of $50\%$ below the vacuum-state level. Furthermore, we have found
that the mean photon number of the superposed cavity modes is twice the mean photon number of either of the two cavity modes. 
Moreover, we have established that the superposed cavity modes are in a squeezed state and the squeezing occurs in both quadratures, 
with the product of the uncertainties in the two quadratures satisfying the uncertainty relation. Furthermore, we have noticed that 
the sum of the squeezing in the plus and minus quadratures is equal to the squeezing in cavity mode $a$ or $b$. We anticipate the 
occurence of squeezing in the two quadratures to have some potential applications.

\end{document}